Natalya Rashevs`ka

PhD of Higher Mathematics Department

Krivoy Rog National University

Viktoriia Tkachuk

Assistant of Engineering Pedagogy and Language Training Department

Kryvyi Rih National University

# TECHNOLOGICAL CONDITIONS OF MOBILE LEARNING IN HIGH SCHOOL

*This paper reviews the history of mobile* learning, *provides a definition of «mobile* learning». *The properties, advantages and disadvantages of mobile* learning, *areas of its implementation at the Technical University and mobile learning tools were specified.*

**The problem statement.** Mobile learning is a new educational paradigm which also is a basis of new learning environment where students can get access to learning materials at any time and from any place they like, and which makes the process inclusive and motivates to continuous education and training during the whole life.

**The analysis of the recent studies** has shown that the prerequisites for mobile learning were established in the 70-ies of the last century when Alan Kay proposed the idea of a book size computer for educational purposes. In the 90 years since the pocket computers have been emerged, the introduction of the mobile learning to the studying process began at the universities, and the first training projects appeared for the implementation of the mobile learning. The foreign scientists investigated the mobile learning in their researches: T. Anderson develops the theoretical and methodological principles of e-learning, M. Sharplz and J. Ettevel looked at the impact of the mobile learning; M. Ragus studies the Australian state mobile learning standard; J. Traksler examines the prospects for the mobile learning development [1].

At the same time the native researches that can systematically describe the technology of mobile learning are limited.

**The aim of the article** is the analysis of the modern state of mobile learning and the determination of the conditions of its implementation at the high technical educational institutions.

**The main material.** The process of the mobile learning in the national education system is in its formation stage. Nowadays the following stages of its development are formed, which are based on the availability of the technical means for the mobile learning and the mobile access implementation to educational resources. These steps can be characterized in such way:

The first mention about the mobile learning occurs in John Dewey's work called "Democracy and Education" in 1916, which refers to the spread of communication channels and the mobile society. Speaking about communication, J. Dewey meant not only getting and sending the messages (the informational aspect), but also the experience exchange (the educational aspect). According to Dewey, the communication is the main educational process in which the teacher becomes one of the participants in the learning process.

The second phase of the mobile learning began from the 50-ies of the last century, when Alan Kay had started the project of the computerized learning called Dynabook.

On the third stage (70-80 years of the twentieth century) the theoretical understanding, the formation plan of using the technical tools and the application methodology, and also the ability of using the first local and global network for the portable computers were hold (90th years of the twentieth century).

The fourth stage (nowadays) is characterized by the development of the systems supporting the mobile learning.

There are many definitions of the "mobile learning" in the literature, but the common fact for this studying technology is that the physical connection to the cable network is not mandatory [7]. As noted by S. Semerikov, " the mobile learning can be defined as an approach to the studying, where the mobile learning environment is being formed based on the mobile electronic devices where students can use them as the means of the access to the educational materials found in the Internet, in any place and at any time they like"[5, 119].

V. O. Kuklev considers the mobile learning as an e-learning using the mobile tools, regardless of the time and place, using the special software on the pedagogical basis of the interdisciplinary teaching and modular approaches [2].

The unique properties of the mobile learning are:

- The ability to the simultaneous interaction both with a single student and the whole group;

- The possibility of the dynamical generation of the educational material depending on the location of students, studying context and the usage of the mobile devices;

- The ability to perform some student's educational activities in the discrete time at any place and in any time [5, 153-154];

- The possibility of the mixed studying implementation [4].

N. Payne identified ten elements of mobile learning; the main of them are [3]:

1) Students are ready to use the mobile devices for studying in those cases when they can not use a book or a computer;

2) The mobile learning gives the opportunity to use the free time intervals;

3) The mobile applications should be compact and should activate from the place where the work has been interrupted;

4) The mobile applications should be available online and should be synchronized with the mobile learning tools.

J. Traksler identifies several directions of the mobile learning implementation [10]:

- The technologically oriented mobile learning – separate specific technological innovations included to the educational process to demonstrate the technical advantages and the educational opportunities;

- The mini electronic learning - mobile, wireless and portable technologies which are used for the repeatedly implementation of the solutions and approaches, which are already used with the traditional e-learning tools and the transfer of some e-learning technologies, such as virtual learning environment (VLE), to mobile platforms (MLE);

- The mixed learning - is the studying process where the traditional learning technology is combined with the innovative technologies of electronic, distance and mobile learning in order to create a harmonious combination of theoretical and practical components of the studying process;

- The informal, personalized, mobile situational learning – is the mobile technology with the additional functionality, for example, those depending on the location;

- The mobile Training – are the technologies which are used to improve the performance and efficiency of the mobile workers by providing the material support "just in time" and in the context of their priorities;

- The remote (rural) developing mobile learning – are the mobile technologies used to solve the infrastructure and ecological problems and to support the education where the traditional education technology is ineffective.

According to M. Sharplz the special features of the mobile learning include: common on-line work on the project, moblogging (mobile blogging), personalized learning, and working in groups, online research, and equal access to education [9]. Everything indicated by M. Sharplz shows that mobile learning is a socio-constructivist.

The social constructivism is characterized as an active process based on previously acquired knowledge of constructing mental models of the world and practice, and provides the access to a variety of the reality descriptions that can teach the ways of the constructing knowledge based on the individuality and the unique experience of every student [6 ].

The main purpose of the mobile learning is to improve person's knowledge in the field he/she wants and in any time it is needed.

The main advantages of the mobile learning in comparison with the electronic learning can be the following:

- The opportunity to learn anywhere and at any time;
- Mobile devices are more compact;
- Continuous access to the studying materials;
- High interactive learning;
- An easy use and convenience of the mobile services.

The disadvantages of mobile learning can include:

- The fragmentation of studying: the studying requires concentration and meditation, when students need to move and such situations may distract their attention;
- The lack of student's well-developed skills and self-control of the cognitive activity;
- The small screen size and difficulties with the internet access: mobile devices have smaller screen sizes than traditional PCs, and most of the Web-sites are optimized for screens with the high resolution;
- The high cost of the initial investment in the organization of mobile learning: device investing for each student, wireless network, maintenance etc.

Thanks to the modern technologies of the mobile communication (interaction "student - teacher" is made in high-speed messaging environment) the high level of interactivity is provided with the mobile learning and that is the crucial thing for education.

It is clear that not all mobile devices can be used in the mobile learning. Requirements for an "ideal mobile device for learning" are formulated by M. Sharplz [8]:

- The super portability;

- The individuality and adaptability mobile device for capabilities, knowledge and teaching style, support person-oriented approach, instead of the total work or entertainment;

- The unobtrusiveness, the student should be excited of the learning process, but not of the device;

- Available everywhere for communicating with teachers;

- Adaptability to the context of learning and development of skills and getting knowledge by students;

- Stability, suitability for learning management during the long studying process; the personal resources and knowledge are also available regardless of changes in technology;

- Intuitiveness, ability to use by people without any experience.

**Conclusions**. 1. The mobile learning is the logical and innovation process in the education system, which is defined as a learning technology which uses the mobile devices, communication technology and intelligent user interfaces.

2. The main means of the mobile learning are the hardware (laptops, netbooks, Tablet PC, Pocket PC, mp-3 players, electronic books, mobile phones, and smart phones) and the software (systems support education, mobile PCM, SRS-systems, computer mathematics systems) mobile means.

3. The mobile learning is caused by the conditions and level of development of the modern information and communication technologies, state of the modern education, the student's desire to be an active participant in the learning process and get knowledge anywhere and anytime.

4. The terms of the mobile learning in high technical institutions are:

- Free Internet access availability;

- Spread of the mobile devices among the students;

- Readiness of support mobile learning systems;

- Transfer to a mixed model of learning;

- Development of method of learning systems is based on the mobile technology.